\documentclass{ws-acs}

\begin{document}

\markboth{Jorge Kurchan}
{Dense granular media as athermal glasses.}

\catchline

\title{Dense granular media as athermal glasses.}

\author{\footnotesize Jorge Kurchan }

\address{ \it P.M.M.H. Ecole Sup\'erieure de Physique et Chimie Industrielles, \\
10, rue Vauquelin, 75231 Paris CEDEX 05,  France.}

\maketitle

\pub{Received (received date)}{Revised (revised date)}

\begin{abstract}
I briefly describe how mean-field glass models can be extended to
the case where the bath and friction are  non-thermal.
Solving their  dynamics  one discovers a
temperature with a thermodynamic meaning associated
with  the slow rearrangements,
even though there is no thermodynamic temperature  at  level of fast dynamics.
This temperature can be shown to match the one defined on the basis
of a flat measure over blocked (jammed) configurations.
Numerical checks on realistic systems suggest that these features 
may be  valid in general. 
\end{abstract}

\section{Glasses and Dense Granular Matter}

An ensemble  of many elastic particles of irregular shapes at low 
temperatures
and high densities forms a glass --- that is,  an out of equilibrium
system having a relaxation timescale that grows as the system ages.
Granular matter would be just an example of this,
albeit a  rather special one, in that the thermal kinetic energy 
$\sim k_B T$  per particle is negligible and that the gravity field
plays an unusually important role. What in fact distinguishes granular
matter from a glass at zero temperature and very high
pressure is the non-thermal manner in which energy is supplied to the grains
(vibration, tapping or shearing) and lost by them (inelastic collisions).
It is because of this difference that we refer to the  
granular-matter/glass {\em analogy}, rather than {\em identity}.

This analogy was already described at the experimental level
by  Struik \cite{Struik}, 
who presented settling powders as aging systems on an equal footing
with  glasses, and made more explicit by the Chicago group \cite{JN92,SSid}.
From the theoretical
point of view, there has been a free exchange of ideas and models
from one field to the other. (See Refs.
\cite{E92,CH96,NCH96,CLHE97,NTEFF,NC98,K98,S98,BKLS00,CN00,SA00}
 for some examples.)

We can thus view the conceptual passage from  glasses to dense granular
matter
as divided in two steps. The first  consists of studying glass
models  in contact
with a heat bath of  very low temperature, under a strong gravity field, and
 considering  them from the point of view of  the quantities
that are  measured in granular matter experiments.
 The second step  consists of focusing on which new features
are brought in by the non-thermal agitation  and friction mechanisms.

As far as the compaction dynamics is concerned, the second question is
usually considered less relevant:
Thus, in the models, vibration is often substituded
by a thermal agitation satisfying detailed balance; for 
example in lattice models by letting particles
move upwards with probability $p_{up}$ and downwards with probability
$1-p_{up}$  (a thermal bath with temperature $\propto 
\ln^{-1}[\frac{p_{up}}{1-p_{up}}]$). 
However, if the recent analytical developments in glass theory \cite{CRAS} are
to be applied to granular matter, it is unavoidable to face the
question of the non-thermal nature of the energy exchange mechanism,
as we shall see below.

\section{ Cage and structural temperatures in glasses.}

A dozen years ago,  a family of  models was identified as being 
schematic {\em mean-field} versions  of structural glasses, somewhat
like the  Curie-Weiss model is  for ferromagnets.   
Above a critical temperature, the dynamics of these models is given by
mode-coupling equations, or generalisations  of them. 
These equations predict that the relaxation of all quantities proceeds
in two steps: a rapid one given the movement of particles in
a `cage' formed by its neighbours, and a slow one generated by
the rearrangement of cages: the structural or $\alpha$-relaxations.
As the temperature is lowered, the structural relaxation
time becomes larger and larger, and it diverges at the critical
temperature. (This transition is in fact smeared in real life, a fact that
can be understood within the same framework). 

If the systems are
quenched below the transition temperature,  they fall out of
equilibrium: the structural relaxation time is not constant but
grows with  the (`waiting') time elapsed after the quench, 
a phenomenon known as aging.
Alternatively, one can submit a system below the critical temperature
to  forces that, like shear stress, can do work continuously.
The surprising result in this case is that aging is interrupted
(see Refs. \cite{K98,rheology}):
the structural relaxation time saturates to a driving-force dependent
value. This  rejuvenation effect is known as {\em shear-thinning} or
{\em thixotropy},
 depending on whether it applies in the liquid or the glass phase.

Below the transition temperature, the system is out of equilibrium,
either because it is still aging or because of the external forces in
the driven case. 
An old idea \cite{Tool} in glass physics is to consider that the
structural
degrees of freedom remain at a higher temperature (of the order of the
glass temperature), while  the cage
motion thermalises with the bath.
In order to make this idea sharp,
we can ask what would be the reading  of a
thermometer  coupled to  the glass. One can show \cite{Cukupe}
that this is related to the ratio of fluctuations
and dissipation, as we now describe.

 Consider an observable $A$, with zero
mean and with fluctuations  characterised by their autocorrelation function
$C_A(t_w+t,t_w)= <A(t+t_w)A(t_w)>$. Let us denote $\chi_A(t_w+t,t_w)$
the response $\delta <A(t_w+t)>/\delta h$
to a field $h$ 
conjugate to $A$, acting from $t_w$ to $t_w+t$.

If above the glass temperature
 we  plot   $\chi_A$
versus $C_A$ using $t$ as a parameter,
we obtain a straight line with gradient $-1/T$: the
fluctuation-dissipation theorem.
For a system aging or subjected to nonconservative forces below the 
glass temperature we can still make the same plot, using $t$ 
(and $t_w$, in the aging case) as parameters. 
It turns out that one obtains a line with {\em two} straight tracts:
for values of $C_A,\chi_A$ corresponding to
fast relaxations the gradient is $-1/T$, while
for values corresponding to the structural relaxation the gradient is
a constant $-1/T_{dyn}$. The effective temperature $T_{dyn}$ so
defined is in fact the temperature read in 
 a thermometer coupled to $A$ tuned to respond to the
slow fluctuations \cite{Cukupe}. Most importantly, it is
observable-independent
within each timescale.

These facts were originally found  in the mean-field/mode-coupling
approximation for glassy dynamics, and later verified numerically (at
least within the times, sizes and precision involved) for a host of
 realistic glass models \cite{CRAS,rheology,BB}.

The appearence of a temperature $T_{dyn}$ for the slow degrees of
freedom, immediately suggested a comparison 
 with an idea proposed by
Edwards originally for granular matter \cite{E92,E94}. For a glass
at very low temperatures it can be stated as follows:
as the glass ages and its energy $E(t)$  slowly decreases, the value of
all macroscopic
observables at time $t$ 
can be computed from an ensemble consisting of all blocked
configurations (the local energy  minima) having energy $E(t)$, taken
with {\em equal statistic weights}.
This ensemble immediately leads to the definition of an entropy
$S_{Edw}(E)$
as the logarithm of the number of blocked configurations, and a
temperature $T^{-1}_{Edw}=dS_{Edw}/dE$ \cite{hist}.

Now, for the mean-field/mode coupling models, {\em it turns out  
that $T_{dyn}$ and $T_{Edw}$ coincide}, and, furthermore,
Edwards' ensemble defined above yields the correct values for the
observables out of equilibrium \cite{K98}.
This  has been recently checked for more realistic 
(nonmean-field) models \cite{Brey,BKLS00}.

\section{Structural temperature in (dissipative)
granular matter.}

In order to see what new features are to be found in granular matter,
we start with the mean-field/mode coupling models, modifying them in
two ways: Firstly we allow for frictional forces that are non-linear,
complicated functions of the velocities. Secondly, we drive 
the systems with forces that do not derive from a potential (`shear-like'),
or are strong and  periodic in time (vibration and tapping).
We expect that the mean-field glass models thus modified will be
minimal mean-field granular matter models.

We measure as before correlations and responses, and, in particular 
diffusion \\ $<|x(t+t_w)-x(t_w)|^2>$ and mobility $\delta
|x(t+t_w)|/\delta f$, where $f$ is a force acting from $t_w$ to $t+t_w$.
The vibrated or tapped case has to be measured `stroboscopically': in
order to avoid seeing oscillations we only consider times that correspond
to integer numbers $n,n'$ of cycles
\begin{equation}
 C(t_n,t_{n'})= <x(t_n)x(t_{n'})>
\end{equation}
\begin{equation}
 \chi(t_n,t_n')= \frac{\delta <x(t_n)>}{\delta f}
\end{equation}
where the force acts during an integer number of cycles from $t_{n'}$
to  $t_{n}$.

In the thermal case we found that
above the glass temperature the comparison of correlations and
responses yields the bath's temperature (as it should, in an
equilibrium situation), and below the glass phase in addition a 
temperature $T_{dyn}$ for the slow degrees of freedom.
For the granular athermal case, this already poses a problem, as not
even in a liquid-like fluidised state do we have a well defined
temperature! (In other words, a parametric $\chi_A$ versus $C_A$ plot
 will not give a
straight line independent of the observable $A$).     
This will  also be 
true for the  `cage' motion \cite{Hansen} in the dense regime.

Surprisingly enough, the next step came from the treatment of 
quantum
glasses at zero temperature at the mean-field level.
 It turns out \cite{CL,Chamon} that these systems
obey a quantum fluctuation-dissipation theorem in the cage motion,
but a classical one in the slow, structural motion:
the nature
of the bath is irrelevant (in the sense of renormalisation group)
as far as the slow motion is concerned.
 In the context of granular matter, a similar 
reasoning~\cite{jorge-trieste,tapping} shows that while there is
 no well-defined dynamic temperature associated to the fast
relaxations --- and in the fluidised regime these are the only
relaxations present,  the slow structural
relaxations still satisfy a  fluctuation-dissipation relation,
with an observable-independent temperature $T_{dyn}$ (see figs 1,2).

\begin{figure}[htb]
\centerline{
        \psfig{file=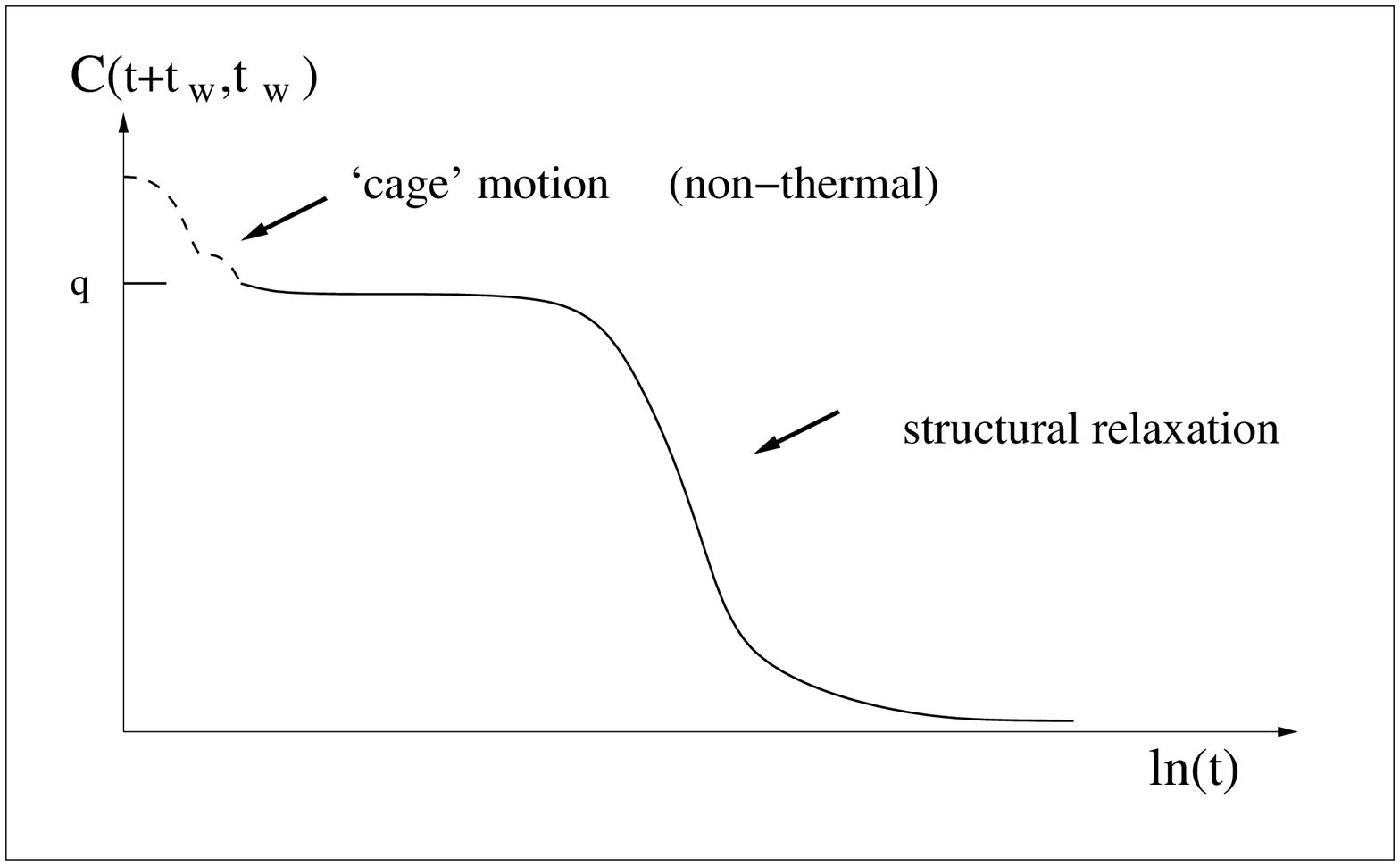,width=6cm,height=6cm,angle=0}
        \psfig{file=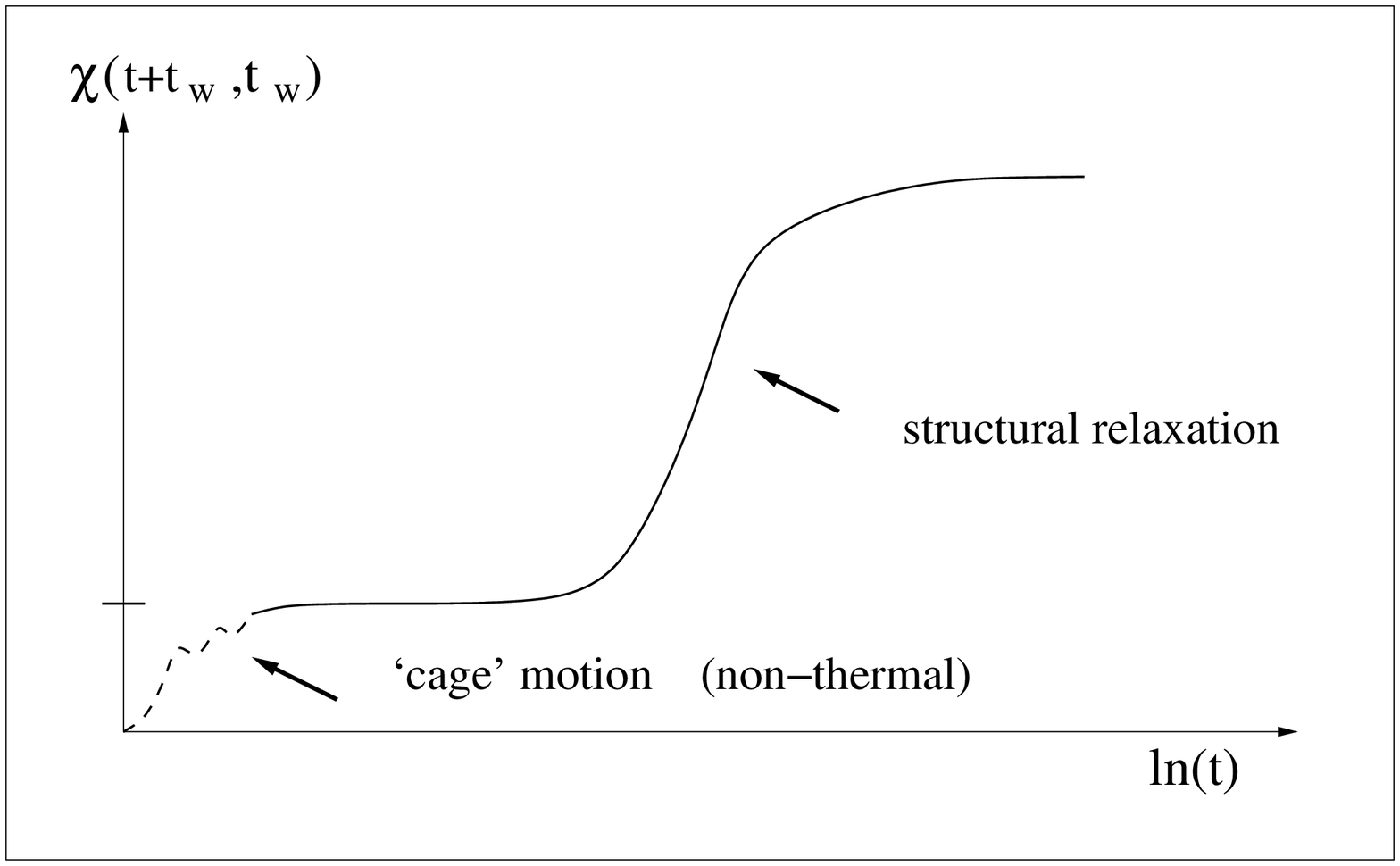,width=6cm,height=6cm,angle=0}}
        \vspace*{8pt}
\caption{Sketch of a  correlation (left) and a response (right) vs. time.}  
\label{twoplots}
\end{figure}

\begin{figure}
\centerline{\psfig{file=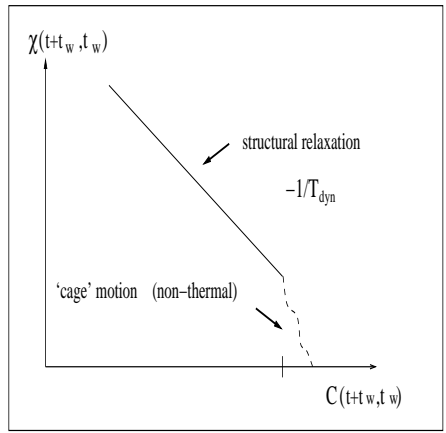,width=6cm,height=6cm}}
\vspace*{8pt}
\caption{Effective temperature  plot. The  dashed
 tract (fast relaxations) is not straight  and  is observable dependent.
The full line (structural relaxations) is straight and defines an
observable-independent temperature.}
\label{coinse}
\end{figure}

 Once these questions 
have been clarified at the level of mean-field/mode-coupling
models, one feels encouraged to check them numerically
and experimentally in realistic systems \cite{Sid2}.
Recently \cite{KM}, a simulation of granular matter subjected to shear
has given evidence for the existence of a
structural temperature. This dynamical temperature is calculated
from the relation between diffusivity and mobility of different
tracers, and its  independency of the tracer shape is checked.
The interest of this setting is that it can  be implemented experimentally.

Within the same model,
a direct computation a thermodynamic temperature defined on the basis of 
the blocked configurations has yielded very good agreement
with the dynamical temperature.

\section{ Conclusion}

 In conclusion, there has been progress in the theory of
 statistical ensembles for dense granular matter.

\begin{itemlist}

 \item We have a better idea of how we should 
  understand them, and of their possible domain of validity.

 \item We have solvable 
 models, and a limit in which we can check if and when 
 these ideas hold strictly.

 \item We have suggestions for experiments that will
 test the validity of the approach in each case.

\end{itemlist}



\begin{thebibliography}{99}
\bibitem{hist} A brief historical note:\\
Edwards himself never defined such a temperature, and indeed
one of the main aims of his  articles cited above is to advocate the
use of volume rather than energy as the relevant variable - leading to
`compactivity' (a kind of pressure) rather than to temperature.

However, in those articles he takes for granted a flat average over 
blocked configurations: this is the aspect of those works  which 
interests us  here. Once an `ergodic' notion is defined, one can choose 
which ensemble is the most appropiate in each case (fixing volume 
and/or energy, etc), and the thermodynamic variables follow.

`Edwards' temperature' here is short for: `the temperature that is
obtained from a flat ensemble \`a la Edwards'.
\bibitem{Sid2} The question was already experimentally adressed in 
Ref.~\cite{SSid}.
The evidence there is not conclusive either for or against
 a thermodynamic approach \cite{KM}.
\end{thebibliography}
\end{document}